# Localized Acoustic-Event Measurement Probe: Connector Confirmation Utilizing Acoustic Signatures

**Brian Skoglind, Travis Roberts, Sourabh Karmakar**

**Clemson University**

Clemson, SC

**Cameron Turner, Laine Mears**

**Clemson University**

Clemson, SC

**ABSTRACT**

Modern consumer products are full of interconnected electrical and electronic modules to fulfill direct and indirect needs. In an automated assembly line still, most of these interconnections are required to be done manually due to the large variety of connector types, connector positions, and the soft, flexible nature of their structures. The manual connection points are the source of partial or completely loose connections. Sometimes connections are missed due to the application of unequal mating forces and natural human fatigue. Subsequently, these defects can lead to unexpected downtime and expensive rework. For successful connection detection, past approaches such as vision verification, Augmented Reality, or circuit parameter-based measurements have shown limited ability to detect the correct connection state. Though most connections emit a specific noise for successful mating, the acoustic-based verification system for electrical connection confirmation has not been extensively researched. The main discouraging reason for such research is the typically low signal-to-noise ratio (SNR) between the sound of a pair of electrical connector mating and the diverse soundscape of the plant. In this study, the authors investigated increasing the SNR between the electrical connector mating sound and the plant soundscape to improve connection success detection by employing a physical system for background noise mitigation and the successful met noise signature amplification algorithm. The theory is that an increase in the SNR will lead to improvement in fault identification through a robust acoustic event detection and classification system. Digital filtering has been used in the past to take care of the low SNRs, however, it also posed the risk of filtering out some potential important signatures for classification. An independent sensor platform has been designed to filter out and reduce background noise from the surrounding assembly line without affecting the raw acoustic signal of the electrical connection, and an automated detection algorithm is presented. The solution is over 75% effective at detecting and classifying connection state. The solution has been constructed without any modification to the existing manual interconnection process.

## 1. Introduction

### 1.1. Problem Description

Electrical and electronic connections in consumer products, even in an automated assembly line, are typically made manually due to the large variety of connector types and the physical flexibility of the designs. The open-loop assembly and quality check for such operations is a "push-pull-push" instruction given to the operator. In such a procedure, first PUSH the connectors together, then PULL them apart to test the locking feature, and lastly PUSH the connector together again to confirm proper seating. This approach is prone to defects as it relies on human subjective interpretation.

Connections can fail by different means and mechanisms [1]. Some of major failure reasons are incomplete insertion due to low force applied, resistance by contaminants at the insertion points, incorrect alignment of connector male and female halves though alignment features are normally included in the design, incorrect manufacture of locking mechanisms (*e.g.*, flashing or short shot). The prescribed quality check method is both physically stressful to the operator due to ergonomic issues caused by the position of the connectors and inadequate to fully verify connection security due to high variation in forces among operators, and by time of shift, and skipping the verification check. An efficient and reliable sensing method is needed to assist operators in confirming successful connections and alleviating the variation in this process.

Acoustic event detection and classification (AEDC) of an electrical connection in an assembly environment is a two-part problem. The first requirement is isolating and detecting the sound of an electrical connection from the loud and diverse soundscape of an assembly plant, followed by signal processing to determine if a proper connection has happened or not. This work presents the design of a mechanical sensor platform to isolate and filter the acoustic signature of the connection event, and a classification approach for determining the connection quality.

The present section of this document serves as an introduction. The second section reviews some of the past works done by researchers to mitigate the connection quality issues through different methods. Section 3 reviews the details of the hardware designed to capture the specific noise signature emitted by the connectors during successful mating. Section 4 covers the use of this developed system in experiments and data collections from a running assembly line. Section 5 focuses on discussions on the results obtained from the data collected from the assembly line. Finally, section 6 provides a conclusion.

### 1.2. Background

In a manufacturing or assembly environment, several noise sources dominate over the main required signal making acoustic quality classification difficult. The soundscapes of these environments are generally complex due to the participation of many sources: propagation through air, propagation through structures, diffraction at the machinery boundaries, and reflections from the ceiling, floor, machinery, and the general non-gaussian nature of the noise spectrum [2]. Due to this complex soundscape, traditional means of AEDC became less effective as they have not been able to extract the noise signal of engagement of the electrical connectors out of the background noise. Digital filtering and amplification method have been applied but generally are found to be inadequate when dealing with acoustic event detection other than speech [3]. It is also a common observation that detection of an audio event is more complex than the classification [4].

Past approaches such as vision verification [5] or Augmented Reality [6] have shown limited ability to predict proper connection state. In [7] the authors analyzed the mating process of plug-in cable connectors in robotic wiring harness assembly. They investigated the challenging difficulties for successful connection detection. Their proposed solution by a new gripper design concepts and also reviewed the error detection improvements. Wen-Chiao *et al.* [8] added an inexpensive secondary connector with the main connector to detect disconnection or loose-connection of the main connector utilizing a canary-based approach. In their work, Phillip, *et al.* [9] used a miniaturized ultrawideband (UWB) source and a mini-spectrum analyzer to measure resonant frequency shifts in connector 'S parameters' as a small and low-cost alternative to a large and expensive network analyzer. They defined 'S parameters' as the voltage coefficients of a general electrical network. Rui Ji, *et al*. [10] researched the connector assembly failures in the electrical circuit of a wireless network. They considered the connector failures based on the samples, a MMSE (Minimum Mean Square Error) algorithm and a neural network algorithm were adopted to classify the fault modes. Çağatay Tokgöz *et al*. [11] analyzed a partially inserted electrical connector model based on a two-level nonlinear least squares (NLS) approach. In their model they found that resonant frequency shifts as insertion depth changes from a fully inserted to a barely touching contact by varying only two length parameters. In their another research, [12] they analyzed the electrical connector faults for avionics systems consisting of coaxial cables. They performed Method of Moments (MoM) analyses using commercial electromagnetic simulation software, FEKO, for transverse electric and magnetic (TEM) wave propagation through a connector for detecting the fully inserted to barely touching contacts by measuring impedance. Sathya *et al*. [13] utilized digital filtering methods such as smoothing, adaptive filtering and linear filtering for background noise mitigation in speech signal processing. These methods were shown to reduce ambient noise power by at least 20 dB with minimum speech distortion. However, these methods took advantage of the quasi-periodic nature of speech waveforms whereas this research focuses on detecting very short click events. Joshi *et al*. [14] investigated a methodology to separate the sound of an electrical connection from the background noise using spectral subtraction. Their approach utilized two microphones attached to the associate and showed significant reduction in noise while distinguishing the signal of interest. While this solution was proven effective, it is not a passive solution which was a requirement from the stakeholders.

The present work is done on an electrical connector for a washing machine pressure sensor. This sensor detects the full water condition. A failed or missed connection is typically not discovered until the final product testing is done at the end of the line. Such failure necessitates partially disassembling the machine consuming high extra time and cost. The connection is performed manually by an associate in the middle of the washing machine assembly line. The successful connection between the connectors emits a distinct click sound. The developed sensor system was installed vertically above the operation station facing the assembly location of the connectors to capture the sound of the connection.

## 2. Initial Problem Investigation
### 2.1. Acoustic Signature Investigation

The initial approach to this problem was to gather sufficient information on the environment, the audible signature of the connector, and performance of several off-the-shelf microphones. The connection event under investigation takes place on an assembly line in a large open manufacturing environment where multiple connection events happen at the same station. This environment was characterized by taking several sound clips with a sensitive binaural microphone at several distances from the targeted connection event. A spectrogram was created from a representative 20 second segment of a recording made from 1 meter away and can be seen in Figure 1. This shows that there is near constant environmental noise between 0 and 5 kHz with frequent broadband events that cover the full 0 to 20 kHz spectrum. These sound clips were later used to simulate the factory environment during laboratory testing.

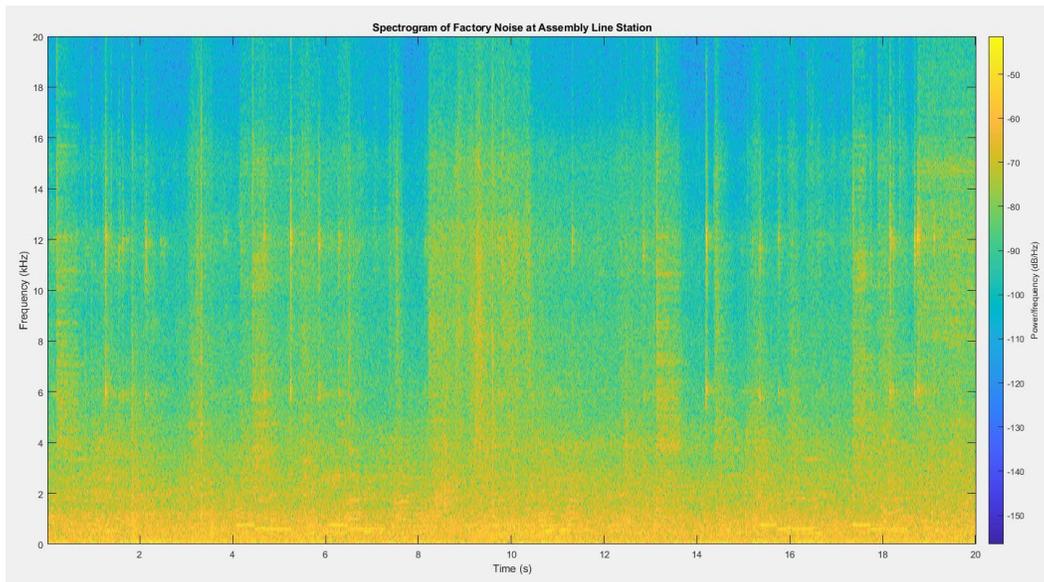

Figure 1: Spectrogram of environmental noise at target facility taken with binaural microphone

The signature of the connection event was characterized by recording the event in a semi-anechoic chamber and analyzing the sound clip in spectrogram format. Figure 2 shows this spectrogram of a single connection event in a semi-anechoic chamber. There is data present across the auditory spectrum (<20 kHz) for about 50 milliseconds, and data in the 1 kHz to 8 kHz band for an additional 300 milliseconds. While equipment was not available for calibrated sound intensity measurements, the factory environment was much louder than the click event.

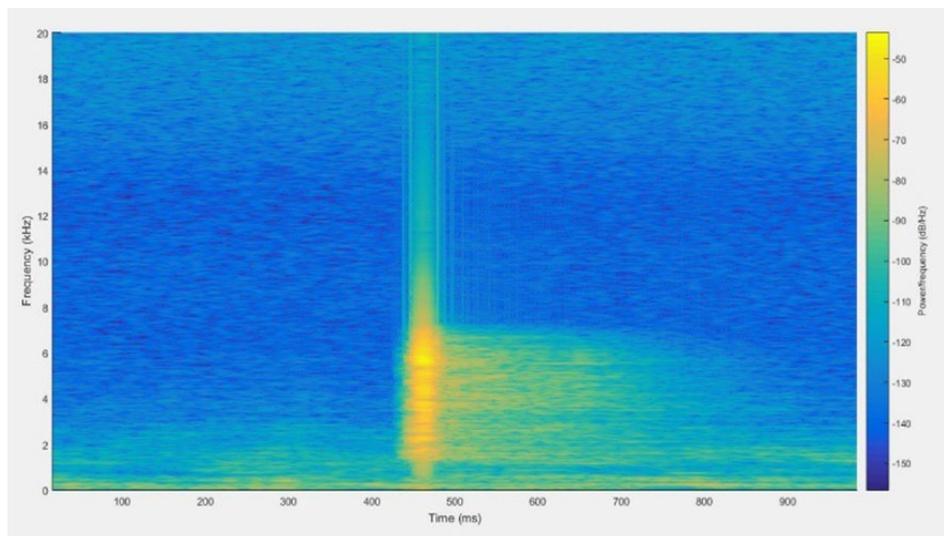

Figure 2: Spectrogram of single connection event recorded in a semi-anechoic chamber.

During the initial investigation, the signature of the click was not detectable in the recordings from the factory as the intensity of the event was below the background noise present in that environment. While digital filtering was initially looked at as a solution for isolating the click signal, simple methods such as bandpass filtering alone would remove significant data from the signal under investigation. Thus, several requirements were developed to delineate the problem: the solution needs to maintain signal fidelity of the click event, increase the intensity of the signal relative to the background noise to a usefully detectable level, cannot interfere with assembly line operations or hinder the worker, and must cost less than $2000.

### *2.2. Initial Solution Testing*

Two devices, a shotgun microphone (Rhode NTG-3 super cardioid) and a parabolic microphone (Sound Shark with Sennheiser ME lavalier microphone), were investigated for their suitability to raise the signal of the click event above the noise. Shotgun microphones are highly directional, picking up sound from directly in front of the microphone much better than other directions. The parabolic microphone uses a parabolic dish to magnify sound along the axis of the dish by reflecting it to the focus point where an omnidirectional lavalier microphone is located. Parabolic dishes are often used for isolating and recording audio in noisy environments such as sports arenas.

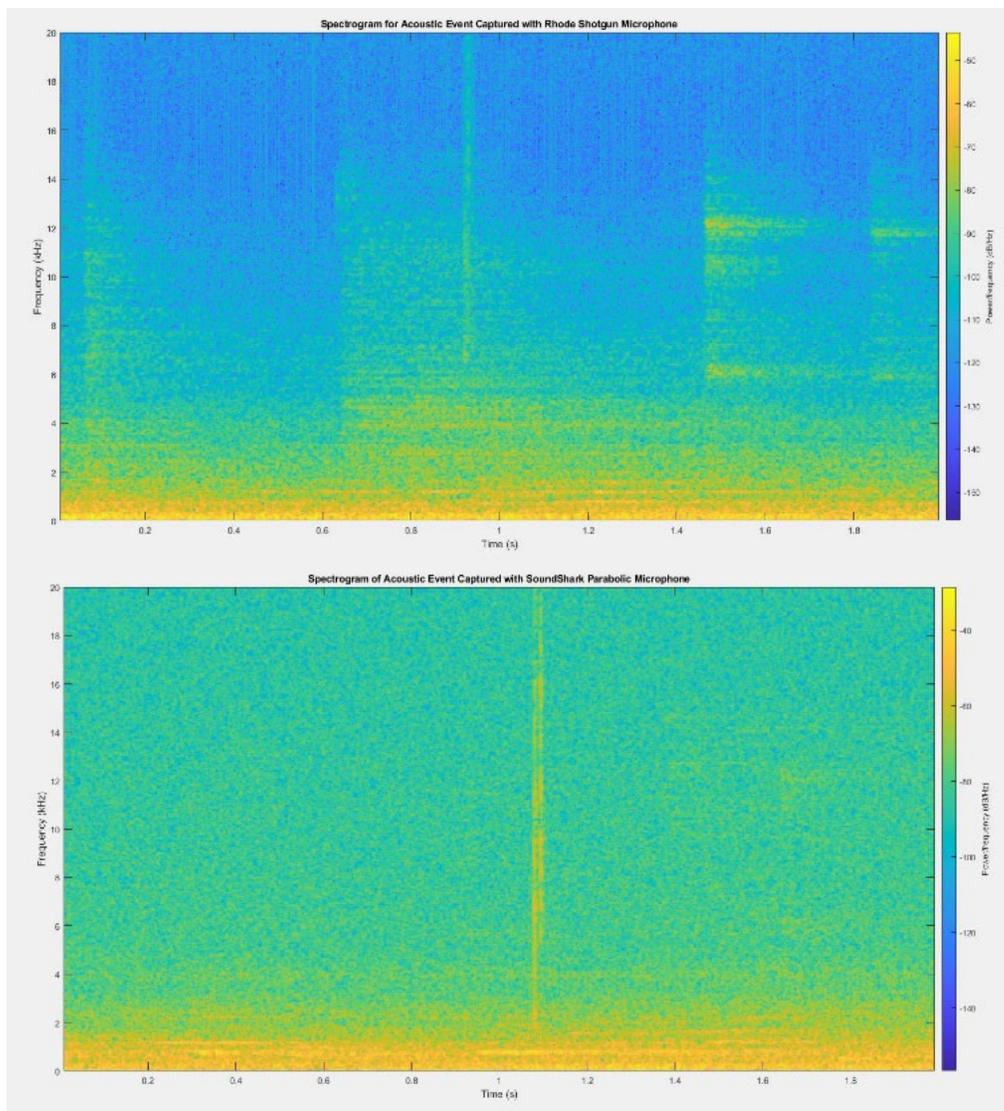

Figure 3: Spectrograms containing a single connection event recorded with (TOP) a Rhode shotgun microphone and (BOTTOM) a Sound Shark parabolic microphone

Spectrograms of 2 second recordings containing a single connection event using both the Sound Shark parabolic dish and the shotgun microphone are shown in Figure 3. The events are distinguishable. However, the shotgun microphone does a much better job of rejecting the background noise than the parabolic dish, but the signal is still not very strong compared to the strength of the environmental noise. The parabolic microphone has done a better job of amplifying the signal of the connection event. The shotgun microphone does show the short duration broadband information but the event data that occurs below the 8 MHz frequency range is obscured by background noise. Neither of these possible solutions performed adequately at rejecting the background noise to be viable on their own.

### *2.3. Design Requirements*

Since the testing of the common off-the-shelf solutions to the problem showed insufficient background noise rejection relative to the level of the acoustic event, it was determined that a custom designed solution would be needed to successfully capture the acoustic event data. Several requirements were developed to constrain the problem. First, the solution should maintain signal fidelity of the acoustic event and should increase the signal to noise ratio of the event relative to the environment so that the event is detectable. The solution should not interfere with the assembly line or associates working on the assembly line. Additionally, the solution should be able to be integrated into the assembly line at some point in the future, though this was not a priority of this initial research and design. Lastly, the solution should also be relatively low cost which in this case meant under $2000. With these requirements in mind, an acoustic sensor platform was developed that was capable of adequately detecting the acoustic event. This sensor was dubbed the Localized Acoustic Event Measurement Probe, or LAMP.

## 3. LAMP Sensor

### *3.1. Prototype Construction*

The proposed sensing platform was constructed around a 24-inch parabolic dish with a cylindrical shield to reduce background noise. The sensor was suspended in the air directly above the assembly line and positioned so that the connection of the electrical connector (the targeted acoustic event) happened directly beneath the dish.

To create the sensing platform, a 24-inch parabolic dish was attached to a circular plywood backing that was 30 inches in diameter. Rubber washers were used to help reduce the transfer of vibrations from the plywood to the dish. An omnidirectional lavalier microphone (Sennheiser?) was positioned at the focal point of the dish using a 3D printed arm with a rear shield to reduce reception of any noise not reflected from the parabolic dish. The circular plywood frame was ringed with a thin HDPE plastic sheet to create a cylinder that was 30 inches in diameter and 32 inches high. The bottom inside of this cylinder was lined with 3-inch thick sound absorbing foam (McMaster-Carr 9710T31). This cylinder extended a maximum of 24 inches below the rim of the parabolic dish. Figure 4 shows a cross section of the CAD representation of the sensor platform. Figure 5 shows side and bottom views of the completed prototype.

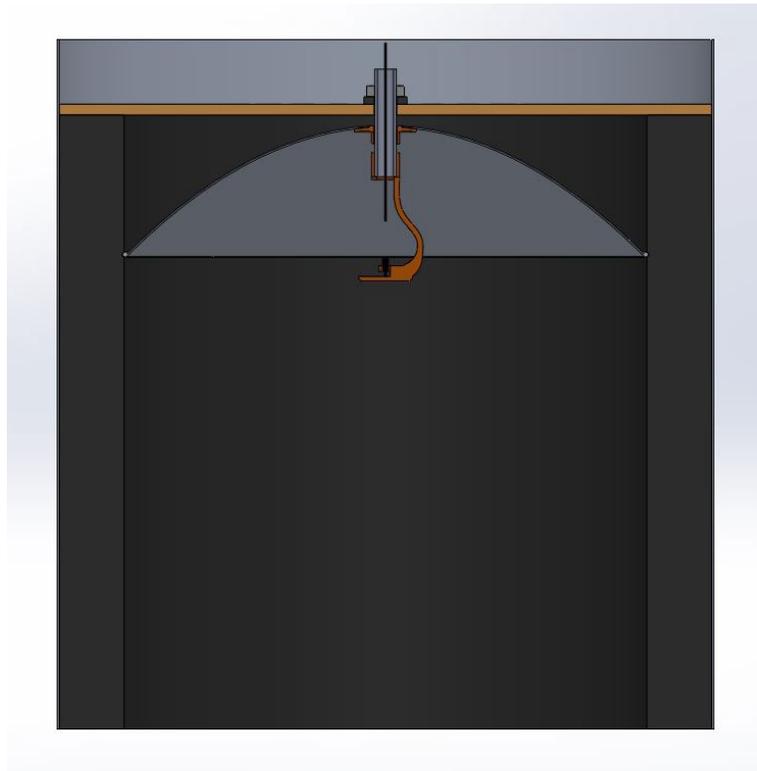

Figure 4: Section view of CAD model for LAMP sensor

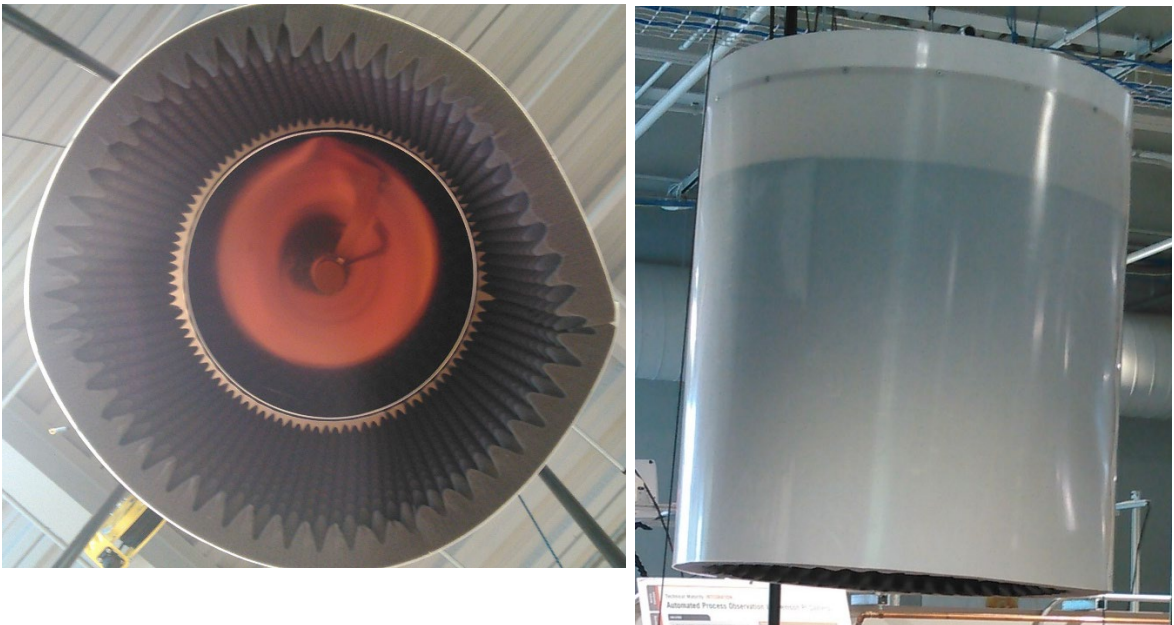

Figure 5: (LEFT) Bottom and (RIGHT) side views of the LAMP sensor

### 3.2. Prototype Characterization

The idea behind the sensor design is that the parabolic dish and the exterior shroud work together to accomplish the design goal; the dish provides amplification of the acoustic event, and the shroud aides in background noise rejection. To determine the effectiveness of the shroud and provide guidance for future prototype revisions, the amount of sound that reached the microphone was measured with the parabolic dish located at differing depths within the exterior shroud. Pink noise was played over speakers in a laboratory setting and recorded using the sensor while suspended at a height of approximately 2 meters above the floor. The speakers were located in the corners of the laboratory so that any noise reaching the sensor would be incidental. The depth of the dish from the bottom edge of the sensor was decreased from 24 inches to a flush mount in 3-inch increments. Each 16 second recording was summed into $1/3^{rd}$ octave power bands for each depth. The results can be viewed in Figure 6 and clearly show that the depth of the shroud beyond the dish has a significant effect on the amount of background noise rejection. The parabolic dish was returned to an inset depth of 24 inches for the rest of the testing.

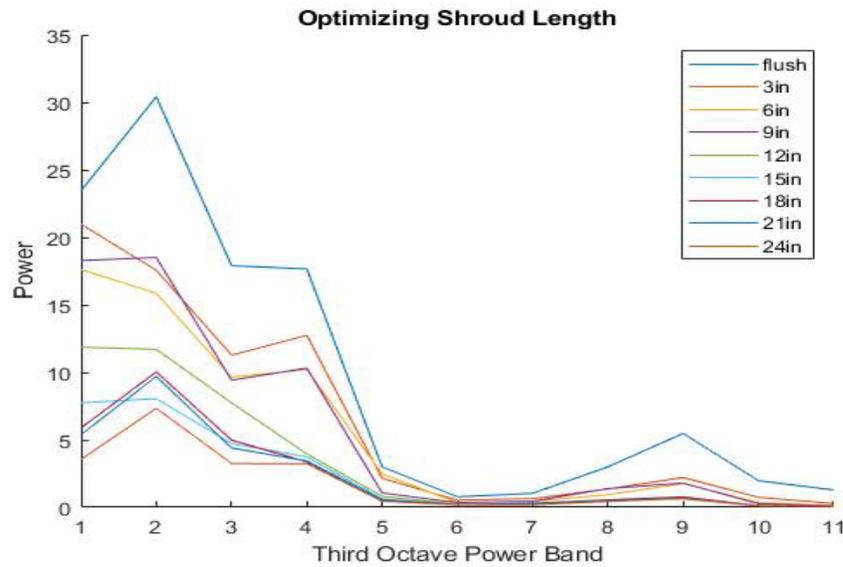

Figure 6: Summed 1/3rd octave power bands for differing amounts of sound shroud extending beyond the rim of the parabolic dish

## 4. Prototype Testing

### 4.1. Laboratory Testing

To simulate the factory environment, a recording of noise taken from the target facility was looped and played over two large speakers in the laboratory environment. The prototype was suspended at a height of about 2 meters from the floor. The authors used the same plastic connectors used in the factory to make a series of connection events about 6 inches from the floor. Figure 7 shows a 2 second sound clip containing a single connection event from the LAMP prototype. This can be compared to Figure 3 to show the improvement in performance of the LAMP sensor over the off the shelf solutions. The shotgun microphone does a decent job of rejecting the background noise but fails to amplify the relatively weak signal of the connection event, but the LAMP spectrogram clearly shows a large reduction in background noise and the connection event signal is intense relative to the level of environmental noise. Figure 8 shows the time data for each of these recordings, and while the connection event is not apparent in the shotgun or parabolic dish signals, it is readily apparent in the LAMP signal. This test proved that the concept of combining the shroud for background noise rejection and the parabolic dish for signal amplification could work to isolate a relatively low magnitude signal in a noise dominant environment, but only factory testing could prove the concept as an acceptable solution to the specific problem being tested

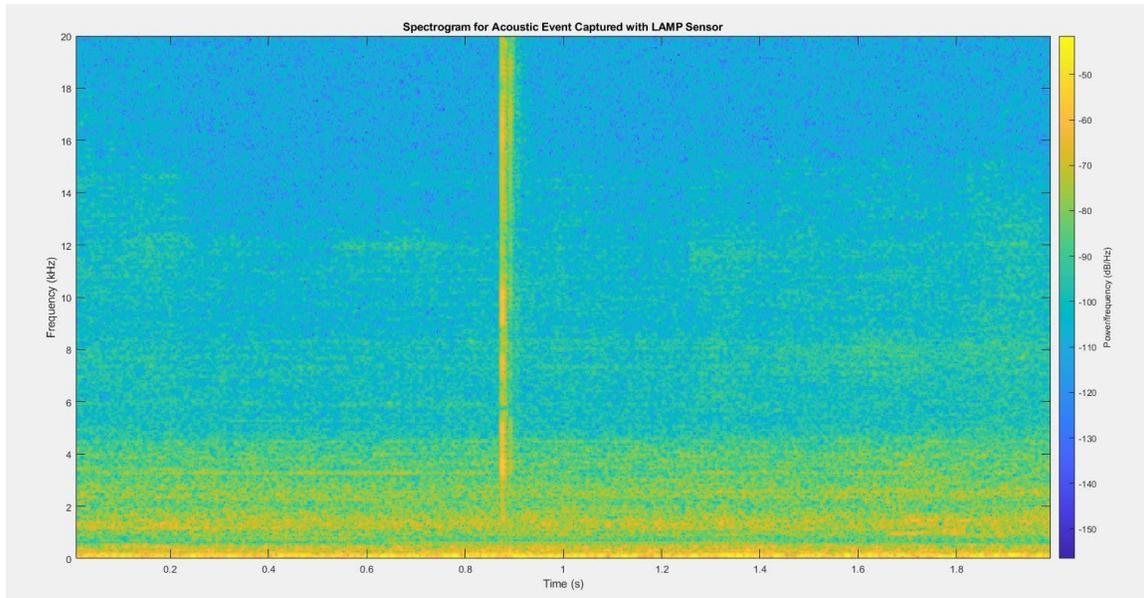

Figure 7: Spectrogram for a single connection event recorded with the LAMP Sensor in lab conditions

## *4.2. Factory Testing*

Returning to the problem requirements, the prototype was suspended at a height of about 2 meters above the location where the targeted connection events take place on the assembly line so as to not interfere with the operation of the assembly line or the movement of the associate. Multiple recordings were taken of normal assembly line operation with simultaneous video recordings to provide ground truth timings for connection events and later analysis. Figure 9 shows a spectrogram recording of a 30 second sound clip with multiple connection events taken in the factory setting. There is significant change in the amount of background noise detected between laboratory and factory environments, but the connection events are still very detectable. This change is attributable to the complexity of the manufacturing environment, which is not easily reproducible in the lab, as well as machinery and parts of the assembly line that were unavoidably in the direct path of the amplification axis for the LAMP sensor. Figure 10 shows a spectrogram of a 2 second sound clip recorded with the LAMP sensor in the factory environment. The factory setting provides a much more challenging environment than the replicated environment in the lab.

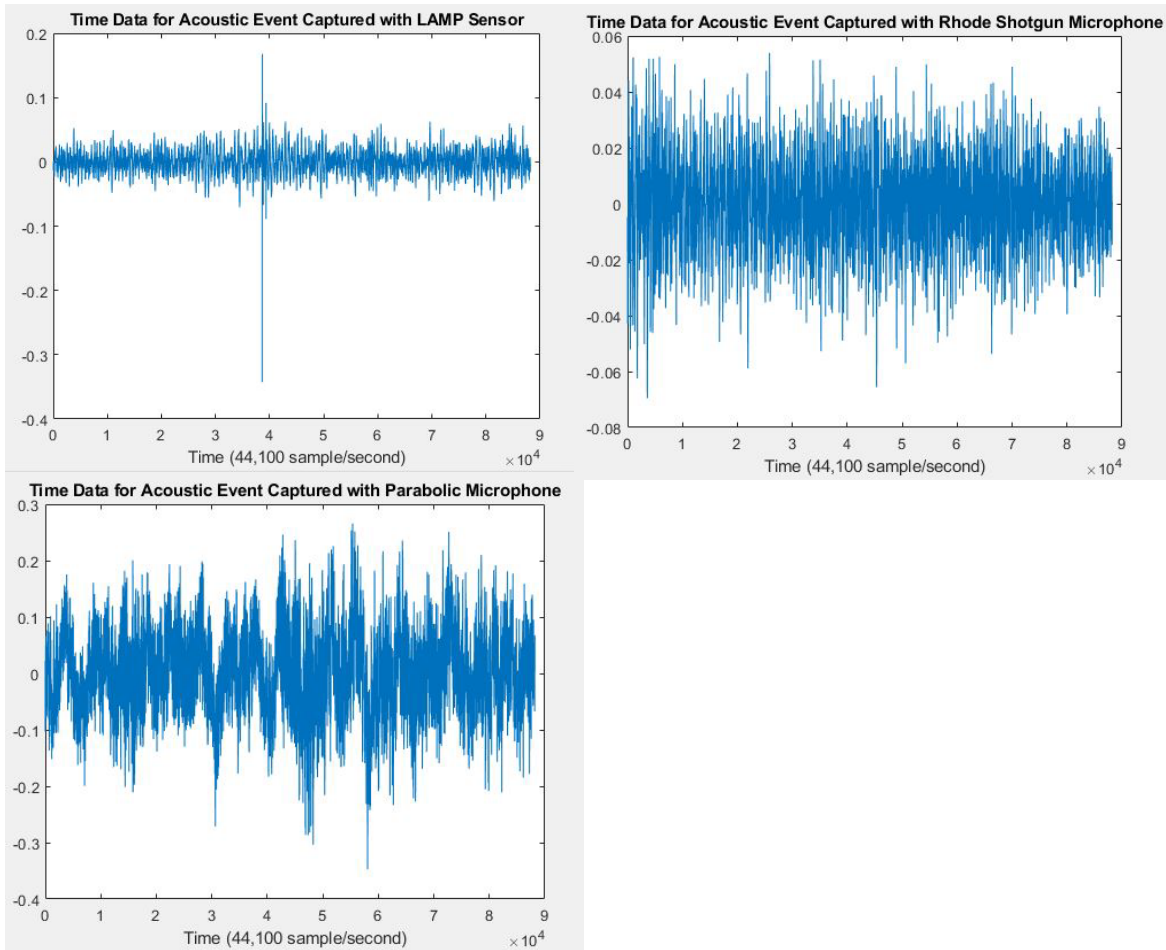

Figure 8: Time data for a single connection event using the (TOP LEFT) LAMP sensor (TOP RIGHT) shotgun microphone and (BOTTOM) parabolic microphone

## 5. Conclusion

### 5.1. Sensor Improvements and Future Work

One of the things that stands out when comparing the spectrograms of the LAMP to the shotgun microphone, is that the shotgun microphone did a good job of rejecting some of the background noise. The LAMP sensor used an omnidirectional lavalier microphone and tried to shield the rear of the microphone with a 3D printed shield in an attempt to only capture sound directly reflected from the parabolic dish. Improvements in future designs may be seen by replacing the omnidirectional microphone with a microphone with a more directed polar pattern such as a cardioid microphone.

This particular prototype was placed a significant distance from the assembly line associate due to its large size and was placed directly overhead to make for easy mounting. A smaller version of the prototype would be able to be placed much closer to the associate without interfering with their work, and thus also be closer to the acoustic event. Additionally, the smaller size would make it easier to mount at a near horizontal angle and a sound absorbing backdrop or panel placed on the opposite side of the assembly station to help reduce any direct reflection of background noise. This could help reduce the environmental noise located along the axis of amplification. Shrinking the size of the prototype could yield gains in mounting flexibility and make the prototype more practical.

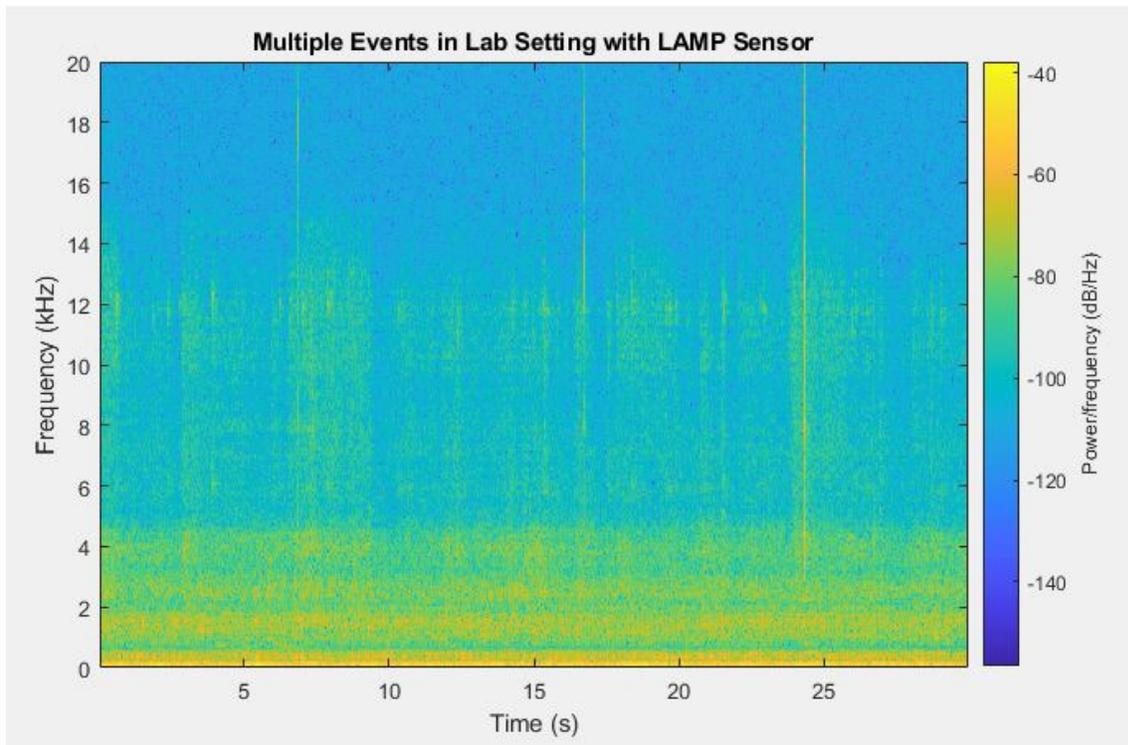

Figure 9: Spectrogram showing three connection events recorded using the LAMP sensor in a lab setting

Comparing Figure 2 with Figures 7 and 10 shows that while the broadband information was able to be isolated using the sound shroud to reject higher frequency environmental noise, there is still significant information from the connection event that could be better analyzed if lower frequency environmental noise could be removed. Pairing the LAMP sensor with some of the software based techniques discussed in Section 1 might make is possible further isolate and extract data from the acoustic event.

This paper has presented the development of an novel acoustic sensor that is capable of magnifying a low amplitude signal in a noisy background through mechanical techniques. The sensor's performance was showcased against several off-the-shelf options for recording acoustic events. Lab and factory testing demonstrated that the sensor is capable of rejecting background noise via mechanical means as well as directional targeting and amplification of a low magnitude signal.

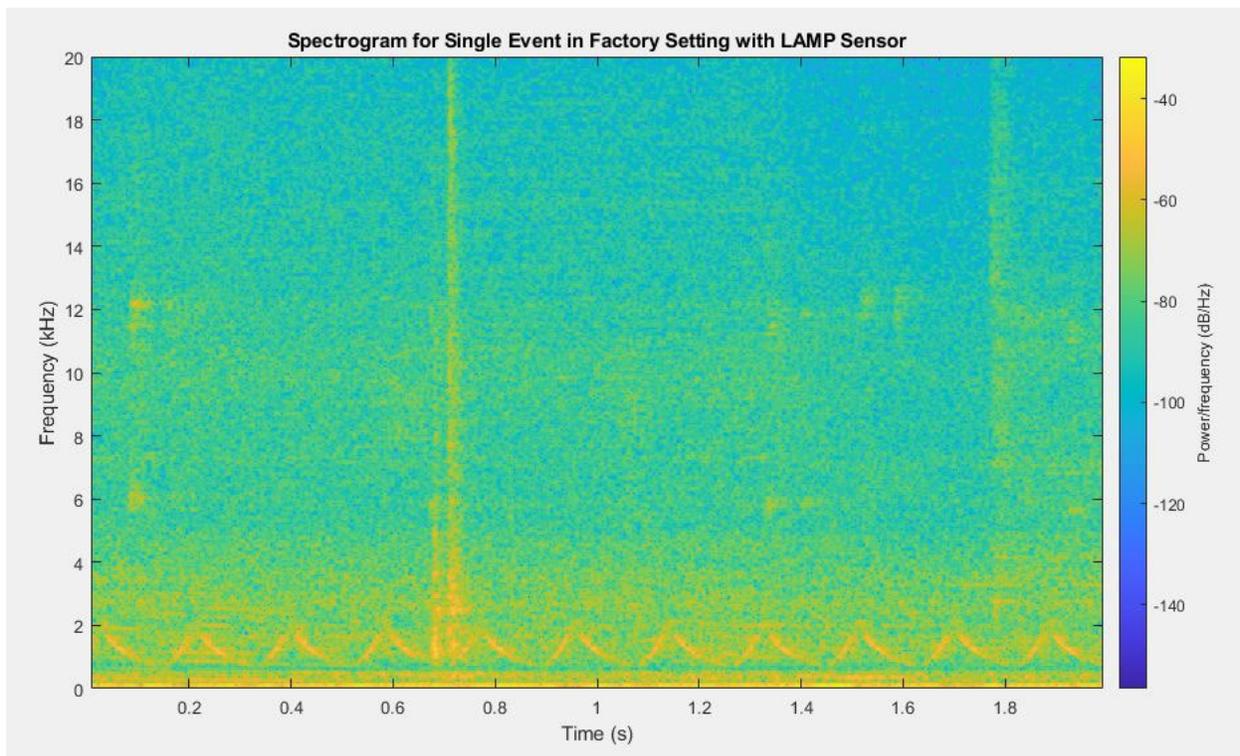

Figure 10: Spectrogram for a single connection event recorded in the factory setting with the LAMP Sensor

## 6. References

[1] B. C. F. de Oliveira, R. C. C. Flesch, A. L. S. Pacheco, and M. B. Demay, "Vision system for detection of defects in the electrical connector of electric motors: test rig and algorithms," *J. Appl. Instrum. Control*, vol. 5, no. 1, pp. 34–42, 2017, [Online]. Available: https://periodicos.utfpr.edu.br/bjic/article/view/5372

[2] R. F. Barron, "Industrial Noise Control and Acoustics," *Ind. Noise Control Acoust.*, pp. 162–224, 2019, doi: 10.1201/9780203910085.

[3] A. Mesaros, T. Heittola, A. Eronen, and T. Virtanen, "Acoustic event detection in real life recordings," *Eur. Signal Process. Conf.*, pp. 1267–1271, 2010.

[4] H. Phan *et al.*, "What makes audio event detection harder than classification?," *25th Eur. Signal Process. Conf. EUSIPCO 2017*, vol. 2017-January, pp. 2739–2743, 2017, doi: 10.23919/EUSIPCO.2017.8081709.

[5] V. Eva, P. Matúš, V. Jozef, O. Stanislav, J. Jozef, and Č. I. Anton, "Detection and classification of audio events in noisy environment," *Doaj*, vol. 3, no. 1, p. 253, 2010.

[6] M. D. Mura, G. Dini, and F. Failli, "An Integrated Environment Based on Augmented Reality and Sensing Device for Manual Assembly Workstations," *Procedia CIRP*, vol. 41, pp. 340–345, 2016, doi: 10.1016/j.procir.2015.12.128.

[7]     F. Yumbla, J. S. Yi, M. Abayebas, and H. Moon, "Analysis of the mating process of plug-in cable connectors for the cable harness assembly task," *Int. Conf. Control. Autom. Syst.*, vol. 2019-Octob, no. July 2020, pp. 1074–1079, 2019, doi: 10.23919/ICCAS47443.2019.8971644.

[8]     W. C. Lin and X. Du, "Prognosis of Power Connector Disconnect and High Resistance Faults," *2018 IEEE Int. Conf. Progn. Heal. Manag. ICPHM 2018*, 2018, doi: 10.1109/ICPHM.2018.8448457.

[9]     Ç. Tokgöz and S. Dardona, "Interrogation of electrical connector faults using miniaturized UWB sources," *Radio Sci.*, vol. 52, no. 1, pp. 94–104, 2017, doi: 10.1002/2016RS006153.

[10]    R. Ji, J. Gao, G. Xie, G. T. Flowers, and C. Chen, "A fault diagnosis method of communication connectors in wireless receiver front-end circuits," *Electr. Contacts, Proc. Annu. Holm Conf. Electr. Contacts*, vol. 2015-Febru, no. February, 2015, doi: 10.1109/HOLM.2014.7031061.

[11]    C. Tokgoz, S. Dardona, and N. C. Soldner, "Modeling of partially inserted connector faults," *2016 IEEE Antennas Propag. Soc. Int. Symp. APSURSI 2016 - Proc.*, pp. 2077–2078, 2016, doi: 10.1109/APS.2016.7696745.

[12]    Ç. gatay Tokgöz and D. Sameh, "Physics-based RF / microwave characterization of wave interactions within electrical connectors," pp. 1489–1502, 2016, doi: 10.1002/2016RS006101.Abstract.

[13]    M. A. J. Sathya and S. P. Victor, "Noise Reduction Techniques and Algorithms For Speech Signal Processing," *Int. J. Sci. Eng. Res.*, vol. 6, no. 1, p. 2015, 2015, [Online]. Available: http://www.ijser.org

[14]    N. S. Joshi, S. Singh, M. Krugh, and L. Mears, "Background noise mitigation of dual microphone system for defect detection in electrical cable connection," *Procedia Manuf.*, vol. 26, pp. 1287–1295, 2018, doi: 10.1016/j.promfg.2018.07.139.